\documentstyle[preprint,eqsecnum,aps,epsf]{revtex}
\newif\iftightenlines\tightenlinesfalse
\tightenlines\tightenlinestrue
\begin{document}
%
\def\eslt{E\llap/_T}
\def\to{\rightarrow}
\def\te{\tilde e}
\def\tl{\tilde l}
\def\ttau{\tilde \tau}
\def\tg{\tilde g}
\def\tnu{\tilde\nu}
\def\tell{\tilde\ell}
\def\tq{\tilde q}
\def\tb{\tilde b}
\def\tt{\tilde t}
\def\tw{\widetilde W}
\def\tz{\widetilde Z}
%
%
\preprint{\vbox{\baselineskip=14pt%
   \rightline{FSU-HEP-940430}\break 
   \rightline{UH-511-788-94}
}}
\title{THE SEARCH FOR TOP SQUARKS AT THE \\
FERMILAB TEVATRON COLLIDER}
\author{Howard Baer$^1$, John Sender$^2$ and Xerxes Tata$^2$}
\address{
$^1$Department of Physics,
Florida State University,
Tallahassee, FL 32306 USA
}
\address{
$^2$Department of Physics and Astronomy,
University of Hawaii,
Honolulu, HI 96822 USA
}
\date{\today}
\maketitle
\begin{abstract}
The lighter superpartner of the top quark ($\tt_1$) may be considerably
lighter than squarks of the first two 
generations, and hence may be accessible to Tevatron collider 
searches, even if the other squarks and the gluino are too heavy.
For the range of $m_{\tt_1}$ of interest at the Tevatron, the 
$\tt_1$ decays to a chargino via $\tt_1 \to b\tw_1$ if this is kinematically 
allowed; otherwise, the flavor-changing loop decay $\tt_1 \to c\tz_1$ dominates.
In the latter case, $\tt_1$ production is signalled by 
jet(s) plus $\eslt$ events. If, instead, the chargino
decay is allowed, $1\ell + b$-$jets + \eslt$ and 
$\ell^+\ell'^- + jet(s) + \eslt$
events from leptonic decays of $\tw_1$ provide the most promising signals.
We perform detailed simulations for each of these signals
using ISAJET 7.07 and devise cuts to enable the extraction of
each of these signals above standard model backgrounds from vector boson or
top quark production. With an integrated luminosity of 100 $pb^{-1}$,
experiments should be able to probe top squark masses up
to $\sim 100$ GeV in each of these channels; the detection of the 
signal in the single lepton channel, however, requires reasonable 
capability to tag displaced $b$ decay vertices in the central region.

\end{abstract}
\medskip
%
%
%
\section{Introduction}

The observation that precision measurements of gauge couplings at 
scale $M_Z$ by the LEP experiments are consistent with the simplest 
supersymmetric\cite{MSSM} SU(5) grand unification\cite{UNIF}
(but not with minimal {\it non}-supersymmetric
SU(5)) has led many authors to seriously re-examine the expectations
for sparticle masses\cite{SUGRA} within this framework. More recently,
several studies of the collider phenomenology
of supergravity models have also appeared\cite{PHEN}.
On the experimental side, direct
constraints on such models come from 
non-observation of super-partners at colliding beam experiments. For 
instance, the non-observation of multi-jet plus missing transverse momentum 
($\eslt$) events above expected
standard model (SM) background levels has led the CDF and D0 collaborations
to conclude\cite{CDF,DZERO},
\begin{equation}
m_{\tg},\ m_{\tq} > 100-150\ {\rm GeV}, 
\eqnum{1}
\end{equation}
where the considerable range in the bound is due to the dependence 
of the missing $\eslt$ cross section on the
parameters\cite{BTW} of the Minimal 
Supersymmetric Model (MSSM). In obtaining these
mass limits, it is usually assumed that there exist ten or twelve
types of mass degenerate squarks. This is usually justified by appealing to 
the framework of minimal supergravity models, where the sfermions
are all
expected to be degenerate at an ultra-high unification scale; this degeneracy
is broken when these $\overline{MS}$ mass parameters are evolved down to the
electroweak scale relevant for phenomenological analyses. For the 
squarks of the first two generations, this evolution is dominantly
governed by their (common) QCD interactions so that their masses
are split only by the relatively small electroweak interactions.
As a result, these remain essentially degenerate with a mass 
$m_{\tq}$. 

In contrast,
the masses of third generation squarks obtain substantial
contributions\cite{STOP} also from the large top family Yukawa interactions,
which also induce a substantial mixing\cite{ER} between the $\tt_L$ and $\tt_R$
squarks. These interactions reduce the diagonal masses relative to
$m_{\tq}$ in much the same way that they drive the Higgs mass squared
to negative values, resulting in the breaking of electroweak symmetry.
At the weak scale, the soft-breaking masses for $\tt$-squarks can be 
written in the form\cite{BOUQUET},
\begin{eqnarray}
& m_{\tt_L}^2 =  m_{\tq}^2 - 2f_t^2m_0^2\Delta_t  - 2f_b^2m_0^2\Delta_b + m_t^2,
 \nonumber \\
& m_{\tt_R}^2 =  m_{\tq}^2 - 4f_t^2m_0^2\Delta_t  + m_t^2. 
\eqnum{2}
\end{eqnarray}
Here, $f_t={{g m_t}\over {\sqrt{2} M_W v}}\sqrt{v^2+v'^2}$ 
($f_b={{g m_b}\over {\sqrt{2} M_W v'}}\sqrt{v^2+v'^2}$) 
are top (bottom) family Yukawa couplings, $m_0$ is the 
universal sfermion mass at the unification scale and $\Delta_{t,b}$ are
dimensionless parameters $\sim 0.1$ that can be numerically computed
within a model. The
term involving the bottom Yukawa coupling is negligible unless 
$\tan\beta (={v\over v'})$
is close to $m_t/m_b$\cite{DN}. Assuming squarks are heavier than
top quarks, we expect the mass ordering $m_{\tt_R} < m_{\tt_L} < m_{\tq}$.
As already noted, $\tt_L$ and $\tt_R$ are mixed by their Yukawa interactions.
The $\tt$-squark mass squared matrix takes the form,
\begin{equation}
{\cal M}^2_{\tt} = \left[
\begin{array}{lr}
m_{\tt_L}^2+m_t^2-M_Z^2\cos 2\beta (-{1\over 2}+{2\over 3}\sin^2\theta_W ) &
-m_t (A_t-\mu\cot\beta ) \\
-m_t (A_t-\mu\cot\beta ) &
m_{\tt_R}^2+m_t^2+M_Z^2\cos 2\beta ({2\over 3}\sin^2\theta_W ) 
\end{array}
\right] ,
\eqnum{3}
\end{equation}
where $A_t$ is the trilinear soft SUSY breaking scalar coupling 
evaluated at the electroweak scale
and $\mu$ is the superpotential Higgs mixing term.
The mass eigenstates ($\tt_1$ and $\tt_2$) can be readily obtained by
diagonalizing this matrix. For the eigenvalues, we have
\begin{eqnarray}
m_{\tt_{1,2}}^2&=& {1\over 2}(m_{\tt_L}^2+m_{\tt_R}^2)+
{1\over 4}M_Z^2\cos 2\beta +m_t^2 \nonumber \\
& & \mp\left\{ \left[ {1\over 2}(m_{\tt_L}^2-m_{\tt_R}^2)+{\cos 2\beta\over 12}
(8M_W^2-5M_Z^2)\right]^2+m_t^2 (A_t-\mu\cot\beta )^2\right\} ^{1\over 2}.
\eqnum{4}
\end{eqnarray}
The corresponding mass eigenstates are
\begin{eqnarray}
\tt_1 =\cos\theta_t \tt_L -\sin\theta_t \tt_R ,\nonumber \\
\tt_2 =\sin\theta_t \tt_L +\cos\theta_t \tt_R ,
\eqnum{5}
\end{eqnarray}
where the mixing angle $\theta_t$ is given by
\begin{equation}
\tan\theta_t={{m_{\tt_L}^2+m_t^2-M_Z^2\cos 2\beta (-{1\over 2}+{2\over 3}
\sin^2\theta_W )-m_{\tt_1}^2} \over {-m_t(A_t-\mu\cot\beta )}}.
\eqnum{6}
\end{equation}

The freedom to adjust the soft-SUSY breaking parameter $A_t$ enables us
to fix the lighter top squark (the stop, or $\tt_1$) 
mass essentially independently of the masses 
of other squarks.
We see that $\tt_1$, which has a mass smaller
than even $m_{\tt_R}$, can indeed be much lighter than the first two
generations of squarks. In fact, $\tt_1$ can easily be as light as $\sim$ 
50 GeV 
even if other squarks and gluinos have masses of several hundred GeV.
Although this is not directly relevant to
our discussion, we also mention that SU(2) gauge invariance
requires that $m_{\tb_L}=m_{\tt_L}$ (aside from $D$-terms) 
whereas $m_{\tb_R}\simeq m_{\tq}$
unless $\tan\beta$ is very large. For moderate values of $\tan\beta$,
$\mu$ and $A_b$, $\tb_L$-$\tb_R$ mixing is expected to be small so
that $\tb_L$ and $\tb_R$ closely approximate the mass eigenstates.

Having convinced ourselves that $\tt_1$ can indeed be much lighter
than other squarks, it is reasonable to ask what current experiments
tell us about its mass. The measurement of the total width of the Z boson
as well as direct searches for squarks at LEP\cite{LEP} imply\cite{FN1} that
\begin{equation}
m_{\tt_1} \agt 45\ {\rm GeV}. \\
\eqnum{7}
\end{equation}
%

%
%
%

Top squarks can also be pair produced at $p\bar p$ colliders, via $q\bar q$
and $gg$ fusion diagrams. Unlike squarks of the first two generations, there
exists no significant cross section contribution from gluino exchange
diagrams, so that $\sigma (\tt_1\bar{\tt_1})$ is determined completely in terms
of $m_{\tt_1}$. For stops with a mass in the 50-125 GeV range--- accessible to 
Tevatron experiments--- the dominant decay mode is expected to be
the two body $\tt_1\to b\tw_1$, if it is kinematically allowed. If instead
$m_{\tt_1}< m_b +m_{\tw_1}$, then decay via flavor changing loop
diagrams is expected to dominate\cite{HK}, in which case $\tt_1\to c\tz_1$.
In the latter case stop production, like squark production (but without
cascade decays), would be 
signalled only by events with jets plus $\eslt$. When
the chargino decay mode is accessible, stop signatures are topologically the
same as those of $t$-quarks. We have checked at least 
within the MSSM framework that if $m_{\tt_1} \alt 150$ GeV,
the decay $\tt_1 \to bW\tz_1$ is kinematically allowed only when the 
two-body decay $\tt_1 \to b\tw_1$ is also allowed. Thus the
three body decay to real $W$ bosons is never phenomenologically 
relevant for the discussion of $\tt$-squark signals at the Tevatron.

The production of $\tt_1$-squark pairs at the Tevatron was studied in Ref.
\cite{BDGGT}, where it was shown that $\sigma (\tt_1\bar{\tt_1})\sim ({1\over
5}-{1\over 10})\sigma (t\bar t)$, for equal top and stop masses. These authors,
using parton level Monte Carlo programs, attempted to translate the
CDF\cite{CDF} squark bound Eq.(1) to a limit on the mass of $\tt_1$, under the
assumption that $\tt_1 \to c \tz_1$. They concluded that the CDF data imply a
stronger bound than the LEP limit in Eq. (4), but only if the lightest
neutralino, also assumed to be the lightest SUSY particle (LSP),
has a mass of just a few GeV. For $m_{\tz_1} > 10$ GeV, 
their analysis showed that it is entirely possible that $\tt_1$
could have escaped detection if its mass was just above the LEP bound.
These authors also studied the dilepton signal from the decay $\tt_1 \to
b\tw_1, \tw_1 \to \ell\nu\tz_1$. Since their purpose was to study whether it
would be identifiable in the CDF search for top quarks, they used the
corresponding CDF cuts. They found that the stop signal would be dwarfed by the
corresponding signal from $t$-quark production. 

Since that time, the CDF experiment has more than quadrupled
its data sample. Furthermore, the D0 experiment has started operation
and accumulated a data sample of comparable magnitude. By the end of the
current run (Run I B), the combined integrated luminosity is expected to be 
$\agt 100$ $pb^{-1}$; {\it i.e.}, $\agt 25$ times larger than that used 
in the earlier analysis\cite{BDGGT}. On the theoretical front, the
production and decay patterns of all sparticles as given by the MSSM
have been incorporated into the event 
simulation program ISAJET\cite{ISAJET}. This should allow for 
considerably better simulation of the signals than in Ref.\cite{BDGGT}.

The purpose of this paper is to re-examine the prospects for discovering 
the lighter top squark at the Tevatron in the case where other squarks 
and gluinos are
too heavy to be kinematically accessible, regardless of whether the
decay $\tt_1 \to b \tw_1$ is allowed. In light of 
the fact that the Tevatron experiments will soon have a very large 
event sample, and considering the improvements in theoretical technology
since the last study, we felt that a reassessment was warranted.
For the $\eslt$
signal we improve on our earlier study in that we use a more sophisticated
simulation. We will see later that the inclusion of QCD radiation in ISAJET
significantly affects our conclusion about the mass reach of the Tevatron,
particularly
when $\tz_1$ is relatively heavy. In the case when the chargino decay
of the stop is kinematically allowed, we adopt a different philosophy
from our earlier study. Instead of focussing on what the Tevatron
top {\it quark} analysis can tell us about top {\it squarks}, we devise
cuts to separate the SUSY signal from the top signal as well as from other
SM backgrounds. We then find that, with an integrated luminosity of
$\sim 100$ $pb^{-1}$, experiments at the Tevatron should be able to search 
for $\tt_1$ as heavy as about 100 GeV, even in this case.

The rest of this paper is organized as follows. We briefly describe our
simulation of stop events at the Tevatron in Sec.~II.
In Sec.~III, we investigate
Tevatron signatures from $\tt_1\to c\tz_1$ decay, and show
regions of the $m_{\tt_1}$ {\it vs.} $m_{\tz_1}$ parameter space
that ought to be accessible to Tevatron collider experiments. In Sec.~IV,
we examine both the single lepton plus jets signature as well as
the dilepton plus jets signature, when $\tt_1\to b\tw_1$ is the
dominant stop decay. Again, we show regions of parameter space explorable via
these modes. We conclude in Sec.~V with a summary of our results along with
some general remarks.

\section{Top squark simulation at the Tevatron}

We use the program ISAJET version 7.07\cite{ISAJET} to simulate events from the
production of $\tt_1$ pairs at the Tevatron. Once produced, the stop rapidly
decays via $\tt_1 \to b\tw_1$ if this decay is kinematically allowed; if not,
we assume that it decays via $c\tz_1$ with a branching fraction of 100\%. 
In other words, the decay patterns of the stop are determined by sparticle
masses
and do not depend on the stop or gaugino mixing angles. 
Furthermore, in this latter
case, the $\eslt$ signal from stop production is completely determined
in terms of $m_{\tt_1}$ and $m_{\tz_1}$ and is independent of any other
parameters. 
If, on the other hand, the stop is heavy enough to decay via the chargino mode, 
its signals obviously depend on the decay patterns of the chargino.
For any set
of input parameters, ISAJET then decays the daughter
chargino with branching fractions as given by the MSSM. Over a large range
of model parameters, the $W^*$-mediated amplitude dominates the decays
of the chargino, so that the branching fraction for the decay 
$\tw_1 \to \mu\nu\tz_1$ is about 11\%. In our computations, we will 
use the supergravity motivated choice $\mu = -m_{\tg}$ as our default
value and fix $\tan\beta=2$. This then implies 
$m_{\tw_1}\simeq 2m_{\tz_1}\simeq {1\over 3}m_{\tg}$,
which together with the stop mass fixes the kinematics of the events.
Effects from radiation off initial and final state partons, fragmentation of
the $c$ or $b$ daughters of the stop, final state hadronization and underlying
event activity are incorporated in ISAJET. 

We have modelled the experimental
conditions at the Tevatron by incorporating a toy calorimeter with segmentation
$\Delta\eta\times\Delta\phi = 0.1\times 0.087$ and extending to $|\eta | = 4$
in our simulation. We have assumed an energy resolution of $50\% /\sqrt{E_T}$
($15\% /\sqrt{E_T}$) for the hadronic (electromagnetic) calorimeter. Jets are
defined to be hadron clusters with $E_T > 15$ GeV within a cone of 
$\Delta R  = \sqrt{\Delta\eta^2 +\Delta\phi^2} = 0.7$ and $|\eta_j| < 3.5$. 
We consider
an electron (muon) to be isolated if $p_T(e) > 8$ GeV ($p_T(\mu) > 5$ GeV)
and the hadronic energy in a cone with $\Delta R = 0.4$ about the lepton
does not exceed $min$($\frac{1}{4}E_T(\ell), 4$ GeV). Non-isolated electrons
are included as part of the accompanying hadron cluster.

\section{Search for stop decay to charm plus neutralino}

When the chargino is heavier than $(m_{\tt_1}-m_b)$, the $\tt_1$
in the mass range of interest essentially decays\cite{HK} via
$\tt_1 \to c\tz_1$ so that stop pair production would be signalled by
events with charm jet(s) together with $\eslt$. In our analysis\cite{FN2}
of this signature, we have imposed the following CDF-inspired\cite{CDF}
cuts on the signal:

\begin{itemize}

\item ({\it i}) We require at least two jets in each event
with at least one jet in the central region, $|\eta_j| \leq 1$.
All jets are required to be separated by at
least $30^{\rm o}$ in azimuth from $\vec{\eslt}$.

\item ({\it ii}) If $n_j = 2$, we further require
$\Delta\phi$($j_1$,$j_2$) $\leq 150^{\rm o}$.
                 
\item ({\it iii}) We veto events containing leptons (from the $c$-jet) with
$p_T$($l$)$\geq 10$ GeV to reduce the background from $W\rightarrow \ell\nu$
($\ell=e$ or $\mu$). We have checked that this leads to very little loss
of signal.

\item ({\it iv}) We require $\eslt \geq 50$ GeV\cite{CDF} to reduce
backgrounds from QCD heavy flavours and mismeasured jets.

\end{itemize}
                  
SM backgrounds from  heavy flavor production ($c\bar c$ and $b\bar b$) 
and multi-jet production (with substantial jet energy 
mismeasurement) are very dependent on a detailed detector simulation, but are
expected\cite{CDF,DZERO} 
to be small, given a large enough $\eslt$ cut and lepton veto. 
The major backgrounds we consider here 
are from vector boson production, and include
({\it a}) $Z\rightarrow\nu\bar{\nu}$ production,
({\it b}) $W\rightarrow\tau\nu$ production (where the hadronically
decaying $\tau$ can be one of the jets), and ({\it c}) $W\rightarrow
\ell\nu$, where extra jets come from initial state QCD radiation. 

The $\eslt$ distribution, before any cuts, from stop events
at a $p\bar{p}$ collider
with $\sqrt{s} = 1.8$ TeV is illustrated in Fig. 1
for two representative choices of $\tt_1$ and $\tz_1$ masses: 
({\it a}) $m_{\tt_1}=85$ GeV, $m_{\tz_1}=20$ GeV, and
({\it b})  $m_{\tt_1}=125$ GeV, $m_{\tz_1}=40$ GeV.
Also shown are the corresponding distribution from the $W$ and $Z$ backgrounds 
listed
above which have been computed using ISAJET. We have used the Set I
structure function of Eichten {\it et. al.}\cite{EHLQ} in our computation.
As expected, the $W \to \ell\nu$ background peaks at $\simeq M_W/2$,
while the $\eslt$ spectra from the other backgrounds are softer. We
see that the $\eslt > 50$ GeV cut ({\it iv}) is very effective in cutting
these backgrounds with a factor of $\sim 2$ loss of signal. Nevertheless, even
after cuts ({\it i-iv}), the $W\to\tau$ ($Z\to\nu\nu$) background is
23 $pb$ (12 $pb$) in contrast to the signal cross section of 9.2 $pb$
and 1.9 $pb$ for our representative cases introduced above. The lepton
veto reduces the $W\to\ell\nu$ background to negligible levels. We
have also checked that $Z\to\tau\bar{\tau}$ is small compared
to the other backgrounds.

To facilitate further separation of the stop signal from SM background, 
we have studied the correlation between  $\Delta\phi$, 
the transverse plane opening
angle between $\vec{\eslt}$ and the nearest jet, and $p_T$(fast jet)
for the $\eslt$ event sample. 
We have illustrated this by scatter plots in Fig. 2 for the 
two signal cases and for the dominant backgrounds. We see that by further
requiring,
\begin{itemize}
\item ({\it v}) $p_T(fast\ jet) > 80$ GeV for $\Delta\phi > 90^{\rm o}$; else
$p_T(fast\ jet) > 50$ GeV,
\end{itemize}
we are able to substantially reduce the main SM backgrounds, with
relatively modest loss of signal (particularly for the $\tt_1 (125\ {\rm GeV})$
case, where the signal was indeed small compared to the backgrounds). 

The resulting cross section from $\tt_1\bar{\tt_1}$ production at
the Tevatron, after cuts ({\it i-v}), is illustrated by
the contour plot in the $m_{\tt_1}\ vs.\ m_{\tz_1}$ plane in Fig. 3. Also
shown are the SM backgrounds after these same cuts. 
We note that although for certain regions of this parameter plane the 
chargino decay mode of the stop will necessarily be kinematically open {\it
within the MSSM framework}, for the purposes of this figure, we have
assumed that the charginos are too heavy to be produced via decays
of the stops, and used the $FORCE$ command in ISAJET to decay the stop
via $\tt_1 \to c\tz_1$. We terminate the contours for stop masses where the
three body decay $\tt_1 \to bW\tz_1$ becomes accessible, since then,
the branching fraction for the two body mode $c\tz_1$ mode is no longer unity.
Several features of Fig. 3 are worth noting. 

\begin{enumerate}
\item We see that the signal cross section after all cuts exceeds 1 $pb$
for $m_{\tt_1}$ as large as $\sim 100-120$ GeV 
even if $\tz_1$ is as heavy as 50 GeV.
For this cross section contour, as many as 
40 stop events may already be present in the accumulated
data sample of the Tevatron. 

\item The dominant SM background source is $W\to\tau\nu$ production, where
the hadronically decaying $\tau$ is counted as one of the jets. Since $\tau$
jets almost always have a charged multiplicity of 1 or 3, it should be 
possible to discriminate the bulk of these events from the signal by vetoing low
charged multiplicity jets.
We show this background to allow the reader
to assess the $\tau$-jet discrimination factor which is necessary for 
sufficient background rejection. If a 
rejection 80\% of $W\to\tau$
events is achieved without substantial loss of signal, 
the stop signal exceeds combined background for $m_{\tt_1} \alt 100$ GeV,
and $m_{\tz_1}<30-40$ GeV.

\item We note that the $Z\to\nu\bar{\nu}$ background can be reliably subtracted 
if a large enough data sample is accumulated for directly
measuring high $p_T$ $Z$ bosons decaying to leptons,
and using the $Z$ branching ratios measured at LEP.

\item Unlike as in Ref.\cite{BDGGT}, where it was concluded that the stop
signal was very small for $m_{\tz_1}\agt 20$ GeV, we see that the signal
is quite robust even if $\tz_1$ is heavy. In fact, we have checked
that there is a significant signal level even if $\tz_1$ is nearly
degenerate with $\tt_1$ --- for instance, for $m_{\tt_1}=m_{\tz_1}+5$ GeV
= 105 GeV, the cross section after all cuts is 0.25 $pb$, which is
more than 2\% of the total stop production cross section of 11 $pb$.
The sharp fall-off of the cross section with increasing $m_{\tz_1}$
in the earlier parton level simulation occured because the daughter charm quarks
rapidly became too soft to satisfy the jet requirements. In the more realistic
ISAJET simulation, 
the stop pair can be produced with a substantial $p_T$ since it can
recoil against jets from QCD radiation. For $m_{\tt_1} \simeq m_{\tz_1}$, the
stops give most of their momentum to the heavy $\tz_1$, so that 
$\eslt \simeq p_T(\tt_1\bar{\tt_1})$ and the jets come from QCD radiation as,
for instance, in the $Z \to \nu\bar{\nu}$ background. 

\item If, as anticipated, the Tevatron experiments accumulate around 
100~$pb^{-1}$ of integrated luminosity by the current run, we may 
anticipate at least 50 or more signal events in the data for 
$m_{\tt_1} \alt 100$ GeV even if $\tz_1$ is relatively heavy.
Although the background
level, as read off from Fig. 3, would be around 200 events (assuming
a $W\to\tau$ rejection of 80\%), we should
keep in mind that the actual background, after $Z \to \nu\bar{\nu}$ events 
are subtracted, will be considerably smaller. A several standard deviation
signal should be possible for stops as heavy as 100-125 GeV. A precise
evaluation of the Tevatron reach, which should take into account 
detector-dependent backgrounds, is beyond the scope of this study.

\end{enumerate}

Up to this point, we have made no use of the fact that the signal always
contains $c$-quark jets. It is clear that SM backgrounds would be considerably
reduced if it were to be possible to tag at least one of the $c$-quarks. This
led us to consider the possibility of using a muon from the semi-leptonic decay
of one of the $c$ quarks as a tag. The signal would then consist of $\mu
+ (n_j \geq 2) + (\eslt \geq 50$ GeV), where the muon is {\it within} a cone of
$\Delta R = 0.4$ about one of the jets. For these muon-tagged
events we require, in addition to ({\it i-iv}) above, that

\begin{itemize}
\item $p_T$($\mu$)$ \geq 3$ GeV for the muon to be identifiable, and

\item either $\Delta\phi$($j_{near}$,$\vec{\eslt}$)$\leq 90^{\rm o}$ or
$p_T$($j_{fast}$)$\geq 50$ GeV. 
\end{itemize}

The signal cross section contours, with these
cuts, are shown in Fig. 4 together with estimates of backgrounds from
$W\rightarrow\tau\nu\rightarrow\mu\nu\nu\nu$,
$W\rightarrow\mu\nu$ and $Z\rightarrow\nu\bar{\nu}+c\bar{c}$ or $b\bar{b}$
processes. To estimate these, we have generated 14K (130K) $W$ ($Z$) events of
each type, and find that just 8, 5 and 6 events, respectively,
pass our cuts. It should also be kept in mind that ISAJET does not include
the full $2 \rightarrow 3$ matrix elements for $Zc\bar{c}$ or $Zb\bar{b}$
production;
in our simulation, these events come from radiation of initial state
gluons followed
by splitting into $b\bar b$ or $c\bar c$ pairs.

We see from Fig. 4 that even with an integrated luminosity of 100 $pb^{-1}$,
there will be only 5-10 tagged events for sparticle masses such that
the $\eslt$ signal shown in Fig. 3
might be difficult to observe above the background. We further see that
the 0.2 $pb$ contour in Fig. 4 (which roughly maps out the region where
$S/B \geq 1$) essentially tracks the contour where the $\eslt$ signal 
in Fig. 3 equals the $Z\to \nu\bar{\nu}$ background. We thus conclude that
the use of muon tagging does not extend the reach of the Tevatron to discover
stop squarks in the $\eslt$ channel. Nevertheless, 
observation of $\mu$-tagged $\eslt$ events could be important,
even at low rates: non-observation of these events at expected rates
would rule out a $\tt_1\to c\tz_1$ decay hypothesis.

\section{Search for stops decaying to charginos plus bottom}

If $m_{\tt_1}> m_{\tw_1}+m_b$, the stop decays via the two-body chargino
mode with a branching fraction of nearly 100\%. Charginos with $m_{\tw_1}\alt 100$ GeV
typically decay via the three-body mode 
$\tw_1 \to f\bar{f}'\tz_1$ ($f =q$ or $\ell$) mediated
by a virtual $W$, $\tq_L$ or $\tell_L$. Since sfermion masses are likely to
be larger than $M_W$, the $W^*$ mediated decays dominate over large 
ranges of SUSY parameters, so that the branching fractions
for chargino decays are, generally speaking, close to those of the $W$-boson. 
It is only when the lightest neutralino is dominantly a $U(1)$ gaugino
that the $W\tw_1\tz_1$ coupling is dynamically suppressed; then 
the sfermion-mediated amplitudes become especially important. If, in
addition, sleptons are also considerably lighter than squarks as is the case
in the no scale limit of supergravity models, the
branching fractions for chargino leptonic decays can be significantly
enhanced. Here, we have mainly confined our attention to the more typical case
where $W^*$-mediated decays indeed dominate. It is, however, worth remembering
that the leptonic signals discussed here may be enhanced in some regions of
parameter space. 

If both the charginos in a stop pair production event decay hadronically,
the signal is qualitatively the same as that in Sec. III: {\it i.e.},
consisting of $n-jet + \eslt$ events. Since, in the present case, 
the $\tz_1$ is always produced via a two-step cascade, we expect\cite{BTW} that 
the $\eslt$ spectrum would be softer than that shown in Fig.~1,
so that SM backgrounds to the signal could be problematic.
For this reason, we focus on the signal where one (or both) of the 
charginos decay leptonically. This stop signal, which consists of
events with 1 or 2 isolated leptons together with jet(s) and $\eslt$ (from
the undetected $\tz_1$'s and neutrino(s)), 
has the same event topology as the canonical signal for top quarks. It
is, therefore, essential to devise strategies to separate the stop signal
events from top background, keeping in mind that cuts designed to
optimize the top quark signal may not be suitable for the detection
of $\tt_1$ squarks\cite{BDGGT}. 

\subsection{The Single Lepton Signal.} 

At the Tevatron, single lepton
events from the cascade decays of gluinos and squarks are expected\cite{BTW} to
be swamped by backgrounds from high $p_T$ $W$ production, with the isolated
lepton coming from the decay of the $W$. What is different about stop
events is that each event contains two hard, central $b$-quark jets which
may be tagged using a silicon microvertex detector.
In our computation, we assume\cite{PRIVATE} 40\% of the events have the vertex
within
the SVX barrel. For these events, we take the efficiency for 
detecting $B$-hadrons with $p_T > 15$ GeV and $|\eta_B| < 1$ to be 30\%. 
In our analysis of the $1\ell$ signal, we have required
the following:
\begin{itemize}
\item ({\it i}) one isolated $e$ ($\mu$) with $p_T > 10$ GeV (5 GeV),

\item ({\it ii}) two or more jets with 
at least one of the jets in the central region $|\eta_j| < 2$,

\item ({\it iii}) $\eslt > 25$ GeV, and

\item ({\it iv}) at least one tagged $B$-hadron.
\end{itemize}
The main SM backgrounds to the single lepton signal 
come from $W \to \ell\nu + jet$ production, and from $t\bar{t}$
production and are shown in Table I for $m_t = 140$ GeV and $m_t = 170$ GeV
along with the signal with $m_{\tt_1} = 100$ GeV, $m_{\tw_1} = 70$ GeV
and $m_{\tz_1} = 30$ GeV. We emphasize that $b$-tagging capability as assumed
above is absolutely essential for the identification
of the signal above the $W$ background. ISAJET generates $Wb\bar b$ events
from initial state gluon radiation followed by gluon splitting, using the usual
collinear QCD matrix elements of the parton shower model.
We have checked that without the $b$-tagging requirement ({\it iv}),
the signal to $W$ background ratio would be degraded by a factor $> 250$.

We see from Table I that the stop signal is considerably 
smaller than the total SM background. We have studied distributions
that may serve
to enhance the signal-to-background
ratio. The multiplicity distribution of jets in the stop
signal (solid) and in the top background for $m_t = 140$ GeV (dotted)
and $m_t = 170$ GeV (dashed-dotted), after cuts ({\it i-iv}) 
is shown in Fig. 5{\it a}. We have
illustrated the signal for $m_{\tt_1} = 100$ GeV, $m_{\tw_1} = 70$ GeV,
and $m_{\tz_1} = 30$ GeV. As expected, the top background has a
significantly larger jet multiplicity than the signal. We have
checked that this is true even if $\tt_1$ is as heavy as 130 GeV.
To enhance the stop signal to top background ratio {\it without much
loss of signal}, we have required 

\begin{itemize}
\item ({\it v}) $n_{jet} \leq 4$. 
\end{itemize}

\noindent In this choice
we have been guided by the fact that the number of signal events
in the data sample of the {\it current} Tevatron run, which is
expected to accumulate an integrated luminosity of 100 $pb^{-1}$,
is not very large. It is clear, however, that if a significantly
larger data sample is available (as should be possible after the
main injector upgrade), choosing $n_{jet} \leq 3$ (recall we always
require at least two jets in the signal) optimizes
the stop signal to top background ratio.

The signal-to-background ratio can 
be further improved by recognizing that in both the top and $W$ backgrounds
the lepton
and much of the $\eslt$ come from the decay of a single $W$-boson. We
thus expect that
the transverse mass distribution $d\sigma/dm_T(\ell,\eslt)$ for the
background exhibits a characteristic Jacobian peak while 
the signal should show no such feature. Toward this end, we
show this distribution, after cuts ({\it i-v}) in Fig. 5{\it b} 
for the same signal case as in Fig. 5{\it a} (solid), the 
$W \to \ell\nu$ background (dashes) and the $t\bar{t}$ background
for $m_t = 140$ GeV (dotted). The transverse mass distribution will thus
exhibit substantial distortion for low $m_T(\ell ,\eslt )$ if a signal 
is present. 
We see that by further requiring,
\begin{itemize}
\item ({\it vi}) $m_T(\ell,\eslt) \leq 45$ GeV,
\end{itemize}
we are able to eliminate a large fraction of the background with a 
relatively modest loss of signal. The background, including cuts ({\it v})
and ({\it vi}),
is shown in the last column in Table I. We see that a signal-to-background 
ratio of
unity is possible for $m_{\tt_1} < 100$ GeV, even for the 
unfavourable $m_t = 140$ GeV case. 
Finally, we examined whether tails 
from direct $b\bar{b}$ production, which has an enormous cross section
at the Tevatron, could possibly fake the stop signal. 
We simulated 3M $b\bar{b}$ events, 
and found that only one
event passes our cuts. Applying the $b$-tagging efficiency,
we estimate that this background to be
$\sim$ 24 $fb$, with large errors.

In Fig. 6, we show contours of constant cross section 
for the single lepton signal in the 
$m_{\tt_1}\ vs.\ m_{\tw_1}$ plane for $\mu = -m_{\tg}$ and $\tan\beta = 2$.
The corresponding $\tz_1$ mass is also shown on the right. 
We have cut off the contours for $m_{\tz_1} < 20$ GeV since this region
is excluded\cite{ROSZ} within the MSSM framework.
We see that unlike
Fig. 3, where the cross section remains substantial even for
$\tz_1$ masses close to $m_{\tt_1}$, the single lepton signal rapidly
decreases when $m_{\tt_1}$ approaches $m_{\tw_1}$. This is because
the $b$-jet becomes very soft and so is not tagged.
We also see that $\agt 20$
signal events (compared to a background of 6-11 events, depending on $m_t$) may
be expected in a Tevatron data sample of 100 $pb^{-1}$ for $m_{\tt_1} < 90$
GeV if $m_{\tw_1}\alt 60$ GeV. This corresponds to about a
$6\sigma$ effect in the data sample expected to be accumulated 
by the end of the current
Tevatron run, {\it provided that sufficient $b$-tagging efficiency is
achieved.} In the longer term, the Tevatron may be able to probe a stop mass
in excess of 100 GeV in this channel, particularly after the main
injector becomes operational. 

\subsection{The Dilepton Signal.} 

Finally, we consider the dilepton signal arising from 
$\tt_1\bar{\tt_1}$ events where both the charginos decay
leptonically. For this signal, we require:
\begin{itemize}
\item ({\it i}) {\it at least} two unlike sign, 
isolated leptons ($e$ or $\mu$), with $p_T(e) > 8$ GeV
and $p_T(\mu) > 5$ GeV, 
\item ({\it ii}) $20^{\rm o} < \Delta\phi(\ell^+,\ell'^-) < 160^{\rm o}$ 
(to reduce Drell-Yan background),
\item ({\it iii}) at least one jet with $|\eta_j| < 2$,
\item ({\it iv}) $\eslt > 25$ GeV.
\end{itemize}

The SM backgrounds to this signal, which dominantly come from $t\bar{t}$
production and $W$ pair production, are shown in Table II after the cuts 
({\it i-iv}). Also shown is the bound on this cross section we get
from a simulation of 10M $b\bar{b}$ events, none of which pass our cuts.
For comparison, we also show the signal for the same case as in
Table I. We see that the signal is smaller than the SM backgrounds
even for the heavier top case. Unlike the case for the single lepton
signal, there is no obvious kinematic variable like the transverse mass that
can be used to enhance the signal relative to the background. We
note, however, that the leptons from the three-body chargino decays 
are expected to be softer than background, which has  
two-body decays of real $W$ bosons.

With this in mind, we examined several distributions (to be discussed
shortly) that may help to distinguish the top squark signal from the 
SM backgrounds. We found that the distribution of the 
scalar sum of the transverse momenta
of the two leptons plus the $\eslt$ in each event, which we refer to as
``bigness'', or $B$,
\begin{equation}
B = |p_T(\ell^+)| + |p_T(\ell'^-)| + |\eslt|,
\eqnum{8}
\end{equation}
provides optimal distinction between the signal and the (dominant) top quark
background. This distribution is shown in Fig. 7{\it a} for the signal
from a 100 GeV stop decaying into a chargino (solid) and for the $t\bar{t}$
background for $m_t = 140$ GeV. As anticipated, the distribution is
considerably harder for the background than for the signal; requiring
\begin{itemize}
\item ({\it v}) $B < 100$ GeV,
\end{itemize}
in addition to cuts ({\it i-iv}), significantly improves the signal to
background ratio. The SM backgrounds, including this cut, are shown in
the last column of Table II. The difference in the signal and background
``bigness'' distributions is illustrated in an alternative fashion in
Fig.~7{\it b}, where we have shown on the horizontal (vertical) axis the signal
(top background) cross section integrated up to the value of $B$ marked on the
curve in this figure. In other 
words, the signal (background) cross section is 80 $fb$ (15 $fb$) if $B <
80$ GeV. It is clear that the best distributions for distinguishing the signal
from background are those where the slope of the corresponding curve remains
small for a large distance after which it turns upward. Furthermore, since we
can read off the value of $B$ for any point along the curve, Fig.~7{\it b}
enables us to easily find an optimal value for the cut. We see that $B < 100$
GeV that we obtained from the histogram in Fig.~7{\it a} is indeed a good
choice for the cut. 

In Fig. 8, we have shown the analogous curves for other combinations
of the lepton momenta and $\eslt$ that may serve to distinguish the 
signal from background for the same cases as in Fig.~7. 
We show ({\it a}) the bigness distribution (again, for ease of comparison),
({\it b}) the $\eslt$ distribution, ({\it c}) the distribution of the scalar
sum, $|p_T(\ell^+)| + |p_T(\ell'^-)|$, and ({\it d}) the $p_T$ distribution of
the hard lepton. From the shapes of these curves, it should be clear that the
``bigness'' distribution is indeed the best one to distinguish between stop and
top production. 

We have illustrated the $p_T$ distributions of the two
leptons in the dilepton signal as well as the corresponding 
$\eslt$ distributions, after cuts ({\it i-v}), in Fig. 9 for ({\it a}) 
$m_{\tt_1} = 100$ GeV, $m_{\tw_1} = 90$ GeV and $m_{\tz_1} = 42$ GeV, 
({\it b}) $m_{\tt_1} = 100$ GeV, $m_{\tw_1} = 60$ GeV and $m_{\tz_1} = 23$ GeV, 
and ({\it c}) $m_{\tt_1} = 70$ GeV, $m_{\tw_1} = 60$ GeV 
and $m_{\tz_1} = 23$ GeV. The fast lepton and
$\eslt$ distributions in cases ({\it a}) and ({\it b}) are considerably harder
than in case ({\it c}) where the stop is rather light and $m_{\tw_1}-m_{\tz_1}$
is small. Also, for a given value of $m_{\tt_1}$, these distributions become
softer with decreasing difference between the chargino and the LSP mass.
The slow lepton $p_T$ distributions are, however, backed up
against the cut. These distributions also
enable the reader to assess the effect on the signal if the cut on the lepton
$p_T$ is further relaxed. Clearly, capability for detection of leptons near
the edge of the $p_T(\ell)$ cut ({\it i}) in events triggered by jets and/or 
$\eslt$ is required in order not to lose too much of the signal. 

In Fig. 10, we show the cross section contours for the dilepton signal after
all the cuts for $m_{\tg} = -\mu$, and $\tan\beta=2$ in the
$m_{\tt_1}-m_{\tw_1}$ plane. The region below the horizontal dashed line is
experimentally excluded\cite{ROSZ} within the MSSM framework, while in the
region above the sloping dashed line, the stop decays via the flavour-changing
loop decay mode analysed in Sec. III. We see that while the signal is rather
small, it exceeds the SM backgrounds for $m_{\tt_1} < 110$ GeV (130 GeV)
for $m_t = 140$ GeV (170 GeV) even if $\tz_1$ is heavy. In a data sample with 
an integrated luminosity of 100 $pb^{-1}$ that should be accumulated by the
end of the current Tevatron run, we may expect $> $ 10
signal events compared to a background of 3-6 events, depending on
the top quark mass, if $m_{\tt_1} < 100$ GeV. Thus the Tevatron
experiments should soon be able to probe top squark masses up to just under 100
GeV (larger, if the top is discovered and found to be heavy) even in the
dilepton channel. This should increase to 120-130 GeV after the main injector
becomes operational and the top turns out to be heavy. 
We remark that unlike as for the single lepton signal, the
detection of the signal in the dilepton channel does not rely on the capability
for tagging displaced vertices. 

We remind the reader that in all our computations up
to this point, we have chosen parameters so that the leptonic branching
fraction for chargino decays is 11\% per lepton family. 
As mentioned above, this may be considerably enhanced\cite{FN3} if sleptons
are lighter than squarks and the $\tz_1$ is mainly a U(1) gaugino. For instance, for
$m_{\tq} = m_{\tg} = 250$ GeV, $\mu = -400$ GeV with $\tan\beta$ = 20
(which yields $m_{\tw_1} = 70$ GeV, $m_{\tz_1} = 35$ GeV and slepton masses
around 125 GeV), 
this branching fraction increases to about 30\%. For this case, we
find the $b$-tagged single-lepton
(dilepton) cross section after all cuts of 220 $fb$ (780 $fb$) for
a stop mass of 100 GeV to be compared with 140 $fb$ (130 $fb$) in Table I.
Our point here is to illustrate that the leptonic
cross sections can indeed be significantly enhanced for reasonable 
choices of SUSY parameters, and that these enhancements are in keeping 
with expectations from the increase in the leptonic branching fraction
of the chargino. In such scenarios, it may be possible to probe stop
masses significantly beyond 100 GeV by the end of Run I B of the Tevatron.
To illustrate, we have checked that for $m_{\tt_1} = 130$ GeV, the choice
$m_{\tg} = m_{\tq} = 315$ GeV, $\mu = -500$ GeV and $\tan\beta$ = 20,
yields cross sections of 63 $fb$ and 210 $fb$ in the single-lepton
and dilepton channels, respectively. It is amusing to see that the 
dilepton signal can be considerably larger than the signal in the 
single-lepton channel.

\section{Discussion and conclusions}

We have seen that, in the framework of supergravity models, the 
lighter of the two top squarks ($\tt_1$) may be
very light even if the other squarks and gluinos have masses
of several hundred GeV. 
Current Tevatron experimental analyses\cite{CDF,DZERO} on squark masses are 
not applicable to a single light top squark, so that currently the best limit
is still $m_{\tt_1}\agt 45$ GeV, based on non-observation of a signal at LEP.
Hence, one of most favorable routes towards the discovery of supersymmetry at
the Fermilab Tevatron collider may be to search for the light top squark.
We have shown in this paper that Tevatron experiments ought to be able
to explore top squark masses of $\sim 100$ GeV, given an integrated 
luminosity of $\sim 100$ $pb^{-1}$. Thus, Tevatron collider experiments
ought to be able to either discover supersymmetry via the top squark, or place
an important new limit on its mass. Such a limit can serve to constrain
the GUT scale parameter space of supergravity models, where the $\tt_1$
mass is driven to small values by a large top quark Yukawa coupling.
In fact, model builders already have to take care to ensure
that $m_{\tt_1}^2$ is not driven negative, leading to breaking of
electromagnetic and colour gauge symmetries\cite{MSSM}. 

If $m_{\tt_1}<m_b+m_{\tw_1}$, then $\tt_1\to c\tz_1$, so 
collider experiments ought to search for
$p\bar p\to\tt_1\bar{\tt_1}\to c\bar{c}\tz_1\tz_1$, leading to 
a multi-jet plus $\eslt$ signature. Even after suitable cuts, significant 
backgrounds from $W$ and $Z$ production remain. 
The $\eslt$ signal cross section, which depends only on the $\tt_1$ 
and $\tz_1$ masses,
is conveniently summarized by cross section contours in the 
$m_{\tt_1}-m_{\tz_1}$ plane in
Figs. 3 and 4.
Assuming that
the backgrounds can indeed be reliably suppressed/subtracted out 
from the data sample (via $\tau$-jet veto, $Z$+jets total 
normalization, {\it etc.}), we see that $\tt_1$ masses of $\sim 80-100$ GeV
ought to be accessible to Tevatron experiments by the end of the current run. 

If $m_{\tt_1}>m_b+m_{\tw_1}$, then $\tt_1\to b\tw_1$, so collider 
experiments ought to search for 
$p\bar p\to\tt_1\bar{\tt_1}\to b\bar{b}\tw_1\bar{\tw_1}$, where 
$\tw_1\to \ell\nu\tz_1$ or $q\bar{q}'\tz_1$. Just as for the top quark search,
signals appear most promising in the single and dilepton final states. 
For single lepton final states, $B$-tagging capability is essential to
veto the large $W$+multi-jet background. 
The results of our computation of the single-lepton and the dilepton
signal are summarized in Fig. 6 and Fig. 10, respectively.
Given sufficient $B$-tagging capability,
$\tt_1$ masses to 80-100 GeV ought to be explorable at the Tevatron via
the single lepton channel.
For the dilepton signal, major backgrounds come from $t\bar t$ and $WW$
production. Requiring ``soft'' dilepton events, ({\it e.g.}, requiring
$B=|p_T(\ell_1)|+|p_T(\ell_2)|+\eslt <100$ GeV), allows sufficient
rejection of background that, as before, stop masses of $\sim 100$ GeV
ought to be probed at the Tevatron by the end of the current run.

The analysis in the present paper is a conservative one, in that
an additional source of stop squark production, $t\to\tt_1\tz_1$, has been
neglected. This decay mode, which can have branching fractions of up
to $\sim 30\%$, would contribute an additional signal, as well as
diminish the SM top quark background\cite{BDGGT}. It would be
interesting to investigate whether the observation of a top quark
signal at a rate consistent with SM expectations could be used to
independently constrain $m_{\tt_1}$ at a top factory. We have also
been conservative in that for the leptonic signals from stop
production shown in Figs. 6 and 10, 
we have chosen parameters so that the branching
fractions for leptonic decays of charginos are essentially the same as 
those of the $W$ boson; as discussed in the text, the single lepton (dilepton)
signals may be enhanced by a factor of $\sim$1.5 ($\sim$6) for certain ranges of
SUSY parameters.

In summary, the search for stop squarks adds a new avenue to the search for
supersymmetry at the Tevatron collider. Light top squarks have a 
significant degree of theoretical motivation. We have attempted to show 
in this paper that our experimental colleagues ought to be able either
to discover top squarks at the Tevatron, or to significantly extend the 
limit on their mass.

%
\acknowledgments

We thank S. Blessing, A. Boehnlein, A. Goshaw and A. White for discussions.
This research was supported in part by the U.~S. Department of Energy
under contract number DE-FG05-87ER40319 and
DE-AM03-76SF00235. In addition, the work of HB was supported by the
TNRLC SSC Fellowship program. 
%
%
%
%

%
\newpage
%
%

\begin{table}
\caption[]{Cross sections in $fb$ at the Tevatron after cuts
described in the text for SM background to the $B$-tagged single lepton
signal. Also shown is the signal cross section for a representative
case near the edge of the Tevatron reach, where we have taken $m_{\tt_1} = 100$ 
GeV, $m_{\tw_1}=70$ GeV and $m_{\tz_1} = 30$ GeV. Results have summed over $e$
and $\mu$.} 
\bigskip
\begin{tabular}{lrrrrrrr}
process & Cuts ({\it i-iv}) & cuts({\it i-vi}) \\
\tableline
$W \to \ell\nu$ & 440 & 30 \\
$t\bar{t}(140)$ & 450 & 81 \\
$t\bar{t}(170)$ & 170 & 32 \\
$b\bar{b}$ &$24$ &24  \\
$\tt_1\bar{\tt_1}(100)$ & 240 & 140 \\
\end{tabular}
\end{table}
\begin{table}
\caption[]{Cross sections in $fb$ at the Tevatron after cuts
described in the text for SM backgrounds to the dilepton
signal from top squark production. 
Also shown is the signal cross section for the same case as in Table I.
Results have summed over $e$ and $\mu$.}
\bigskip
\begin{tabular}{lrrrrrrr}
process & Cuts ({\it i-iv}) & cuts({\it i-v}) \\
\tableline
$W^+W^-$ & 50 & 10 \\
$t\bar{t}(140)$ &360 &54 \\
$t\bar{t}(170)$ &135 &14 \\
$b\bar{b}$ &$<48$ &   \\
$\tt_1\bar{\tt_1}(100)$ &180 & 130 \\
\end{tabular}
\end{table}


%
\begin{figure}
\caption[]{$\eslt$ distributions, before any cuts, from 
$\tt_1\bar{\tt_1}$ production
in $p\bar{p}$ collisions at
$\sqrt{s}=1.8$~TeV for two choices of stop and $\tz_1$ masses.
Here, the stop is assumed to decay via $\tt_1 \to c \tz_1$.
Also shown are SM background distributions from $Z \to \nu\bar{\nu}$ 
production (dotted), $W \to \tau\nu$ production (dashed) and $W \to \mu\nu$
production (dot-dashed).}
\end{figure}
%
\begin{figure}
\caption[]{The scatter plot for $\eslt$ events in the  
$\Delta\phi(\vec{\eslt},j_{near})$ - $p_T(j_{fast})$ 
plane after cuts ({\it i-iv}) in Sec.~III
of the text for the $\tt_1\bar{\tt_1}$ signal at the Tevatron for 
({\it a}) $m_{\tt_1} = 85$ GeV, $m_{\tz_1} = 20$ GeV,
({\it b}) $m_{\tt_1} = 125$ GeV, $m_{\tz_1} = 40$ GeV,
and dominant SM backgrounds from 
({\it c}) $W \to \tau\nu$ production, and ({\it d}) $Z \to \nu\bar{\nu}$
production.}
\end{figure}
%
\begin{figure} 
\caption[]{Cross section (in $pb$) contours for the $\eslt$ distributions,
after cuts ({\it i-v}) in Sec. III of the text, from $\tt_1\bar{\tt_1}$
production in $p\bar{p}$ collisions at $\sqrt{s}=1.8$~TeV, assuming that the
stops decay via $\tt_1 \to c \tz_1$. Also shown are dominant backgrounds from
SM processes. The $W \to \tau$ background will be smaller if it is possible to
veto $\tau$ jets at some level. Also, it may be possible to subtract the $Z
\to \nu\bar{\nu}$ background as discussed in the text.} 
\end{figure}
%
\begin{figure} 
\caption[]{The same as Fig. 3 except that, in this figure, we require
a non-isolated muon with $p_T(\mu) > 3$ GeV to tag the $c$-jet. Cuts are
as described in Sec. III of the text.}
\end{figure}
%
\begin{figure} 
\caption[]{({\it a}) The jet multiplicity distribution in
$B$-tagged single lepton events from stop pair production at a $p\bar{p}$
collider at $\sqrt{s}=1.8$ TeV, assuming that stop decays via $\tt_1 \to
b\tw_1$ after cuts ({\it i-iv}) in Sec. IV.A of the text.  We illustrate the
distribution from the signal (solid) for $m_{\tt_1} = 100$ GeV, $m_{\tw_1} =
70$ GeV and $m_{\tz_1} = 30$ GeV, and from top quark backgrounds for 
$m_t = 140$ GeV (dotted) and $m_t = 170$ GeV (dashed-dotted), and ({\it b})
the transverse mass $m_T(\ell,\eslt)$ distribution in these events
after cuts ({\it i-v}) in Sec. IV.A of the text. We illustrate the
distribution from the signal (solid) for the same SUSY masses
as in ({\it a}), and from the dominant SM backgrounds from
$W \to \ell\nu$ production (dashed) and $t\bar{t}$ production for $m_t = 140$
GeV (dotted). The branching fraction for $\tw_1 \to e\nu\tz_1$ decay is 11\%.
If the top quark has a mass of 170 GeV, its contribution to the background is
significantly smaller.} 
\end{figure}
%
\begin{figure}
\caption[]{Cross section (in $fb$) contours in the $m_{\tt_1}-m_{\tw_1}$ plane  
for  $B$-tagged single
lepton events from stop pair production at a $p\bar{p}$ collider at
$\sqrt{s}=1.8$ TeV, assuming that stop decays via $\tt_1 \to b\tw_1$
after cuts ({\it i-vi}) in Sec. IV.A of the text.
The branching
fraction for $\tw_1 \to e\nu\tz_1$ decay is 11\%. Also shown are the
main SM backgrounds. The mass of $\tz_1$ is illustrated on the right.}
\end{figure}
%
\begin{figure}
\caption[]{({\it a}) Distributions of the ``bigness'' variable $B$ defined
in Sec. IV.B of the text for dilepton events from stop pair production 
at the Tevatron (where the stop decays via $\tt_1\to b\tw_1$) and from the
background from production of 140 GeV $t\bar{t}$ pairs after cuts ({\it
i-iv}) in Sec. IV.B of the text. The sparticle masses and chargino branching
fractions are as in Fig.~5. ({\it b}) The cross section for stop dilepton
signal (top background) on the horizontal (vertical) axis, integrated up to the
value of $B$ on the curve after the same cuts as in ({\it a}). This plot is
described in more detail in Sec. IV.B.} 
\end{figure}
%
\begin{figure} 
\caption[]{Cross sections with ({\it a}) the bigness, ({\it b}) $\eslt$,
({\it c}) $|p_T(\ell_1)| + |p_T(\ell_2)|$, and ({\it d}) $p_T(\ell_{fast})$
variables integrated up to the values shown on the respective curves, for
stop dilepton events (horizontal axis) and $t\bar{t}$ background, for 
$m_t = 140$ GeV, after cuts ({\it i-iv}) of Sec. IV.B.}
\end{figure}
%
\begin{figure} 
\caption[]{Distributions of the transverse momentum of the two leptons
in dilepton events from stop pair production at the Tevatron along with
the $\eslt$ distribution after cuts ({\it i-v}) in Sec. IV.B for
three illustrative cases of $\tt_1$, $\tw_1$ and $\tz_1$ masses mentioned in 
the text.}
\end{figure}
%
\begin{figure} 
\caption[]{Cross section (in $fb$) contours for the dilepton cross section
from the production of stop pairs at a $p\bar{p}$ collider with $\sqrt{s} =
1.8$ TeV, including cuts ({\it i-v}) of Sec. IV.B. The branching fraction for
$\tw_1 \to e\nu\tz_1$ decay is 11\%. Also shown are the main SM backgrounds.} 
\end{figure}
%


\begin{references}
\bibitem{FN3} In the case where the sneutrino is light enough so that
the two-body decay $\tw_1 \to \tilde{\nu}\ell$ is accessible, the signal
will be greatly degraded since the daughter lepton will generally be soft.
Thus, the enhancement we refer to obtains only when sleptons and sneutrinos
are significantly heavier than charginos, but lighter than squarks.
\end{references}
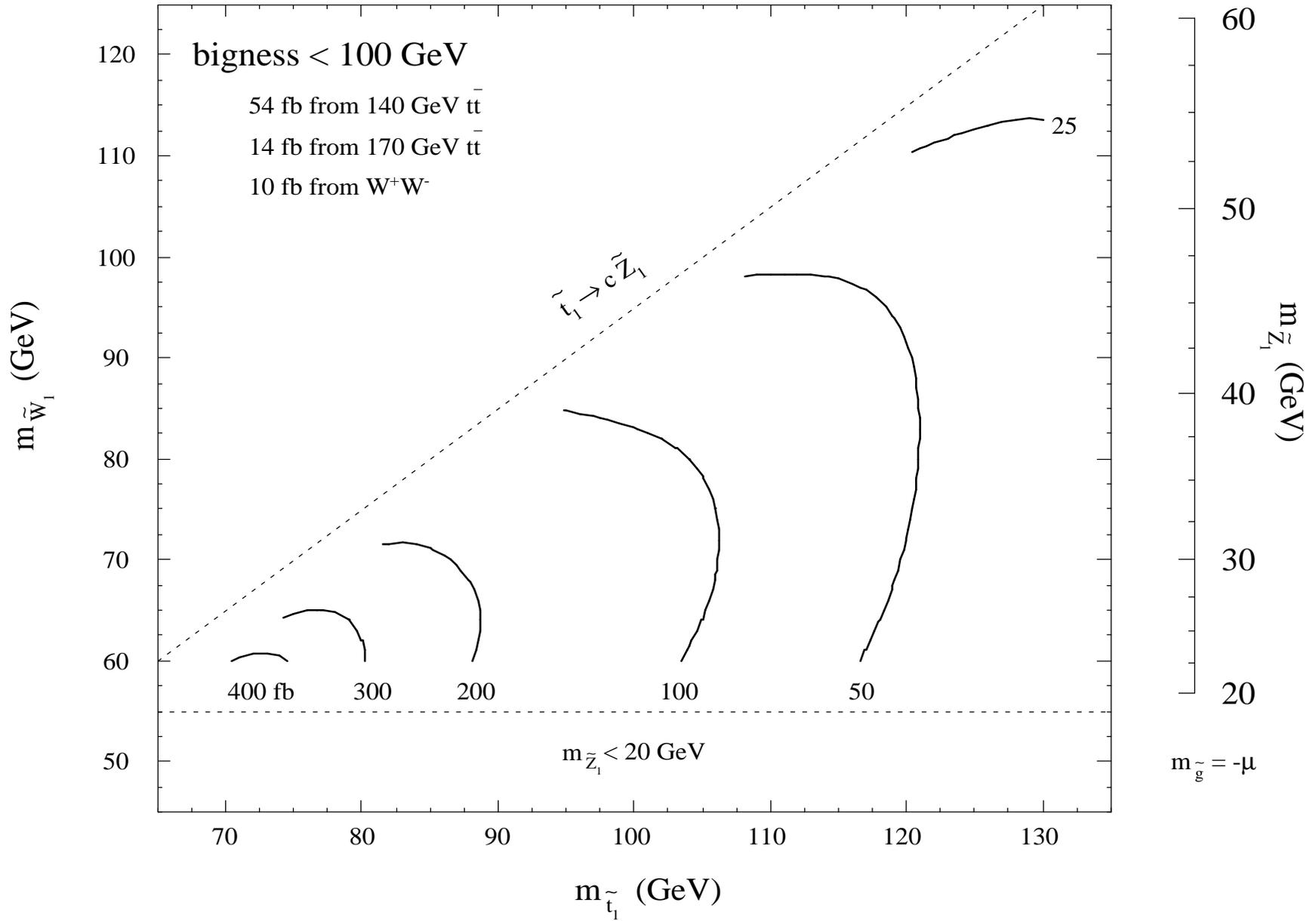
\end{document}